# Pretty Modular Symmetric Encryption (PMSE), compact algorithm for "embedded cryptography" with quite low computational cost

Etienne LEMAIRE

**Université de Lyon, ECAM Lyon, LabECAM, F-69005 Lyon, France**

**Abstract**

Nowadays, the dataflux shared between *IoT* systems must be secured from 8-bits to 64-bits processors systems. Several symmetric cryptographic algorithm already exist such as AES (Advanced Encryption Standard), RC4, Blowfish, etc. In this work, we propose an 8-bits encryption algorithm, combining several ideas of standard symmetric encryption algorithms. The aim is to provide a efficient, modular and compact algorithm able to tend to truly random one-time-pad encryption quality even for large data flux. The algorithm combines the implementation of a divergent polynomial with variable coefficients for pseudo-random keys generation, variable bit swapping and bitwise operations on data, and the use of one or two passwords. The encryption has been evaluated statistically, tested for image encryption and compared with One-Time-Pad encryption on the same data. Three implementations have been tested respectively in C, Javascript and GNU Octave. Encryption time performances are compared with AES on a 8-bits architecture: the Arduino Uno (ATmega328) microcontroller. A new concept of self-decryptionable web encrypted object, called *Blocksnet©* is also presented and tested as a potential application of the compact and embedded PMSE algorithm.

**Keywords:** PMSE, Cryptography, Symmetric, Encryption, Modular, Algorithm, Blocksnet

**Introduction**

The use of encryption in order to achieve authenticated communication in networks have been initiated for long time [1, Needham et al.]. The symmetric cryptography system is a sequence of operations that convert the plain data to encrypted data and the process is reversible in order to get back the original data again [2, Ashraf et al.]. The standard cryptographic algorithm massively used for securing internet flux have some drawbacks [3, Al-Hazaimeh et al.], such as iterative rounds for AES [4, DES et AES], weakness in decryption process for Blowfish [5, Vaudenay et al.], or high calculation cost for public key encryption like RSA (The Rivest, Shamir, and Adleman) [6, Rivest et al, 7, Zhao et al., 8, Rifà-Pous et al.]. Moreover, because no system is truly secured, the worldwide standardization makes one static standard exposed to a large panel of attacks. Therefore, in this work, we build a modular encryption algorithm, that can be tuned at each use for specific cryptographic purposes enhancing security thanks to versioning and a pretty variable algorithmic core.

The PMSE algorithm is based on a dedicated byte stream cipher combining a modular pseudo-random flux and a reversible versionable deconstruction process for data. The pseudo-random flux is calculated from a divergent polynomial of variable coefficients which depends one or two passwords. The order of polynomial is also modulable. Then, a xor cipher is applied between pseudo-random key and the "deconstructed" data byte. The deconstruction of data byte consist of conditional bit swapping where the new arrangement of bits depends on a n-states pseudo-random variable updated at each iteration. The number of possible arrangements for data byte deconstruction is tunable. The reverse operation (called data "reconstruction") does not necessarily use the same operands. This conditional bit swapping combined with variable bitwise operations permits to obtain a non-malleable cryptographic system [9, Fishchlin et al.]. This is an additional security implemented by the PMSE algorithm.

## I. Algorithm of PMSE

PMSE use a modified Vernam system, demonstrated as theoretically secured when taking into account the Shannon's conditions [10, Shannon]. PMSE includes the generation of a flux of pseudo-random bytes, and the "deconstruction" of data byte possibly in one of $N$ different possible arrangements, before encryption. The encryption final step consists of the bitwise *xor* operation between modified data byte and a pseudo-random byte for each byte of the flux. The schematic representation of encryption and decryption is represented in Figure 1. A sequence (e.g. a loop) is initiated with the first byte of data and ended with the last one. Data byte are deconstructed and combined ("*XORed*") with a different pseudo-random key at each iteration. At the end of this loop, the encryption process is done. The General idea of the algorithm is to pseudo-randomize not only one flux, but also the data and pseudo random generator itself. Therefore pseudo-random flux and data deconstruction processes are both highly modulable and versionable.

a)

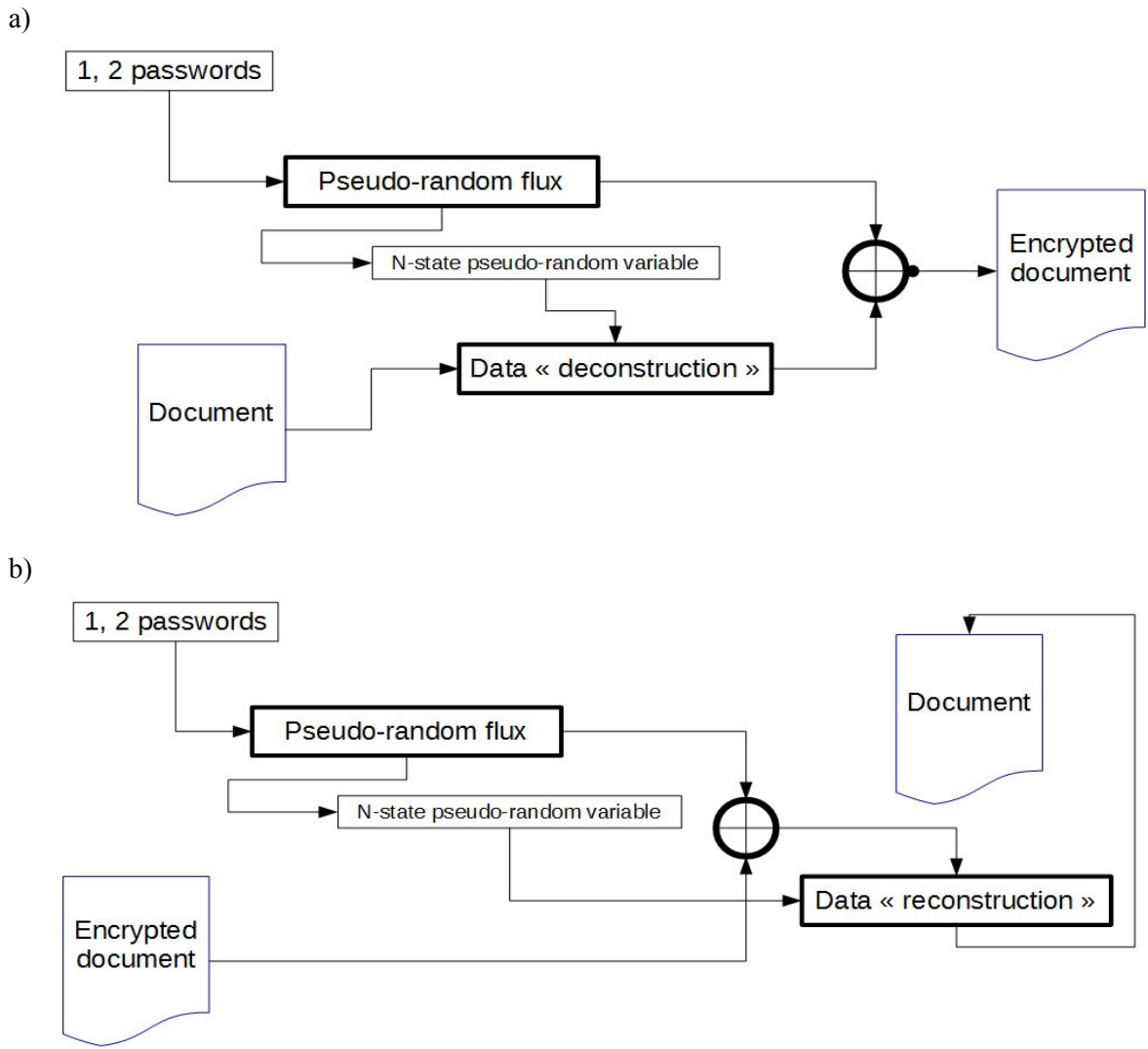

b)

*Fig.1 Schematic representation of the a) encryption and b) decryption flux.*

The pseudo-random flux calculation is described in Figure 2. It consists of the calculation of a polynomial at each iteration. For instance, a simple first order polynomial could be:

$$L_n = x_2.i + x_1 \qquad (Eq.1.1)$$

Or a second order polynomial, including recursivity could be:

$$L_n = x_2.i^2 + x_1.i + L_{n-1} \qquad (Eq.1.2)$$

Where $x_1$ and $x_2$ are one of the characters of respectively first and second passwords (or two concatenated parts of a single password). And $i$ is an iteration number (byte index into the flux).

The general form of a modular recursive polynomial could be:

$$L_n = \sum x_k.i^k + \sum L_{n-m} \qquad (Eq.1.2)$$

Where the $x_k$ are several keys of first and second passwords, $i$ is an iteration number (byte index into the flux) and $L_{n-m}$ the previous values calculated for $L_n$ that are added.

The result $L_n$, (large number: eg. 32 or 64 bits integer) is then cutted into four or respectively 8 bytes, all combined together into a new byte $x0$ with bitwise *xor* operation. At each iteration the value of $x_0$ will change even if the same pair of $\{x_1, x_2\}$ or the same $\{x_k\}$ vector is used, due to $i$ incrementation. Then the pair $\{x_1, x_2\}$ or the $\{x_k\}$ vector is updated from passwords strings. Another temporary key is also calculated from the following equation:

$$x_3 = x_1.i + x_2.x_3 \qquad (Eq.2)$$

The result $x_3$, (32 bit integer), depends on the updated $x_1$ and $x_2$, previous $x_3$ and $i$ the incrementation index. Only the 8 LSB of $x_3$ will be used for the current pseudo-random key calculation. Finally the current pseudo-random key $xt$ is calculated from $x_0$, $x_1$, $x_2$, $x_3$ and previous $x_t$ using bitwise xor operation. This key $x_t$ will be used for the encryption with the current "deconstructed" data byte.

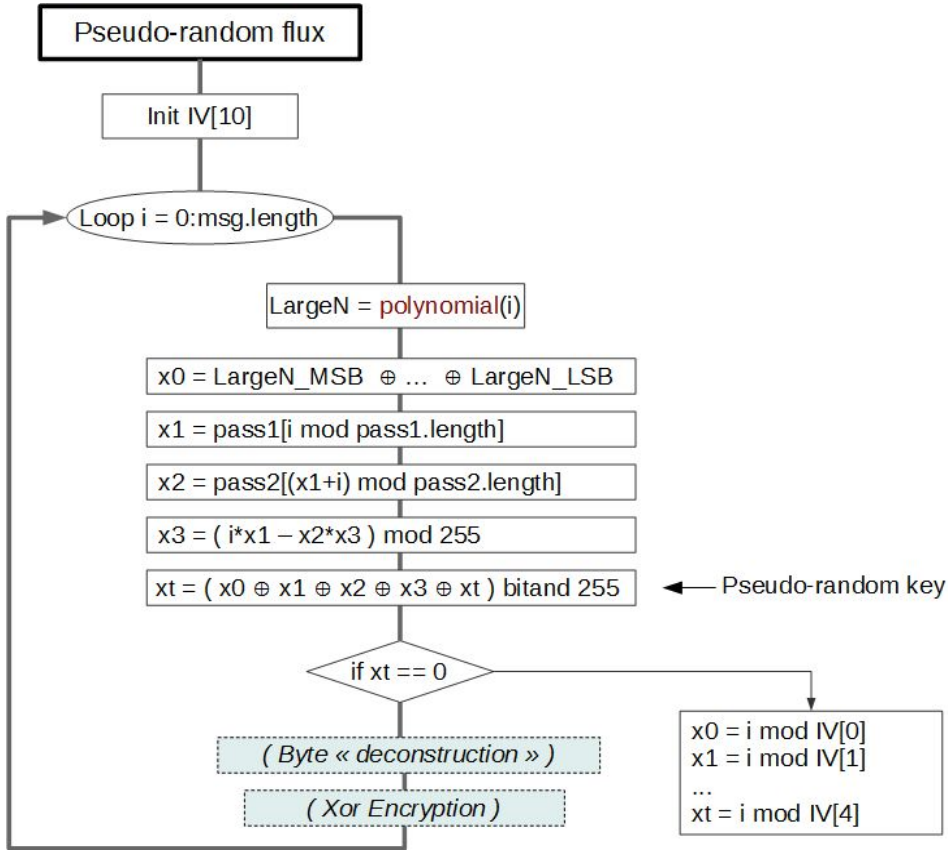

*Fig.2 Graphical representation and pseudo-script of the PMSE pseudo-random key generation algorithm.*

If for some index $i$, the current $x_t$ is equal to zero, the whole calculation is reinitialized using initialization vector ($iv$). As shown is Figure 2, the values $x_0$, $x_1$, $x_2$, $x_3$ and $x_t$ are reinitialized as follow:

$x_0 = i \bmod iv[0]$         $x_1 = i \bmod iv[1]$         $x_2 = i \bmod iv[2]$

$x_3 = i \bmod iv[3]$         $x_t = i \bmod iv[4]$         (Eq.3)

This exception allows to set a non-zero value for $x_t$ depending of iteration index modulo an initialization value. In the same time, for the same iteration, the current data byte is "deconstructed". As illustrated in Figure 3, a bit swapping occurs depending of the value of $x_0$ modulo the number of possible bits swapping and partial complementation implemented. Theoretically, the number of swapping or permutation combination is higher than byte range. But some permutation give the same binary number (only 256 possible values). Additionally the logical *not* operation can be applied to a part or to the whole byte in order to increase the number of available reversible bits operations that can be performed on the current databyte. In practice, implementing the byte deconstruction process with only four cases will already distort significantly the original data (as shown in next part). The implementation of some reversible permutations and/or not operations as well as the number of cases are modulable from a version to another version of the algorithm and as mentioned, a lot of combination of operation can be used. The more permutation cases are implemented, the more robust and secured the PMSE algorithm is. Notice that the maximum number of permutations and/or bitwise operations should be less than 256, due to the fact that the data byte and $x_0$ are both bytes ranging from 0 to 255.

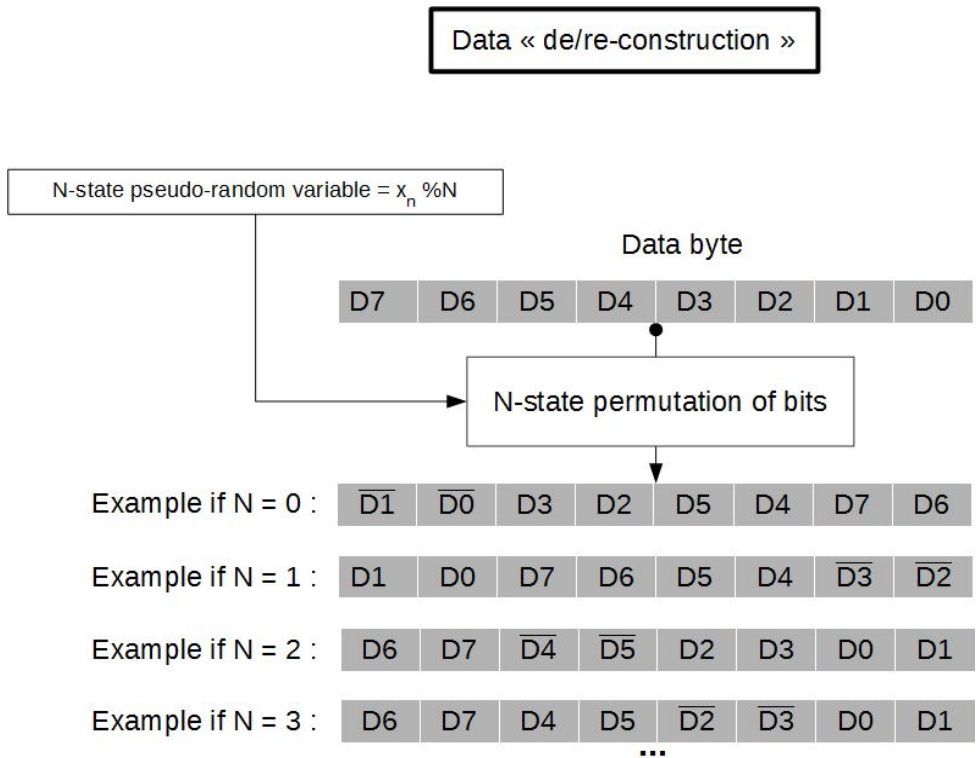

*Fig.3 Illustration of pseudo-random bit swapping of data performed for each data byte for encryption and in the opposite way for decryption.*

## II. Tests and results

### II.a Some entropy tests and encryptions using GNU Octave

The present algorithm has been implemented into GNU Octave (source codes provided into the Supplementary data section) in order to test the entropy of the pseudo-random generator and to perform some image encryption. The generation of 10000 bytes has been performed using two small passwords : 'aa' and 'bb' (Figure 4). The signal observed on large (Figure 4-a) or small (Figure 4-b) intervals doesn't present visually any repetition pattern. The histogram of the distribution of bytes is quite uniform on the [0-255] space. The FFT spectrum of the signal generated (Figure 4-d) with these small passwords also suggests that the signal is close to some experimental noise without any dominant frequency or possible identification in the frequency domain for this size of sample (10 kB).

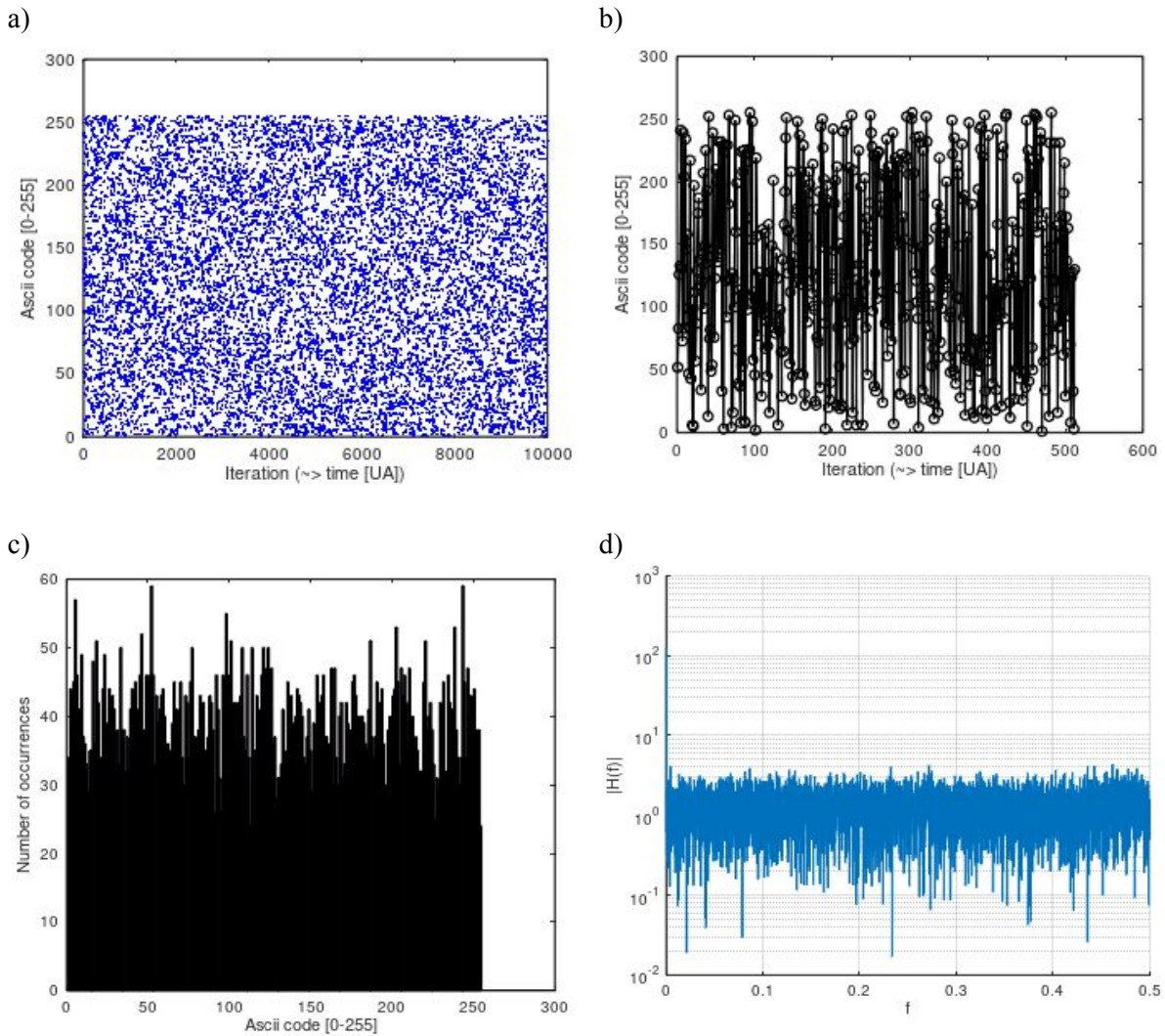

*Fig.4 a) Generation of 10000 pseudo-random bytes using the algorithm with the passwords respectively equals to 'aa' and 'bb'. b) Partial sample of signal of the pseudo-random bytes. c) Histogram, repartition of the 10000 bytes in the byte space [0-255]. d) FFT of signal of 10000 pseudo-random generated bytes (10 kB). The polynomial used is the one described in Eq.1.1, the figures have been obtained using the code given in the supplementary data section.*

In order to complete this first test, some statistics have been computed on several vectors of pseudo-random numbers generated with different sets of small (unsecured) passwords with PMSE. The statistical correlation between different or similar passwords set is also calculated (Table 1). The correlation between vectors of different passwords tends to really low correlation with negative score close to zero. Some vectors calculated with similar sets of password give low correlation score by changing only one single letter ('pass' or' mass'). Moreover the generation of two pseudo-random vectors with the same set of passwords in different order (such as 'abc' and 'bcd' or 'bcd' and 'abc') also give low correlation value.

| **Passwords** | 'aa' 'bb' | 'bonjour' 'hello' | 'abc' 'bcd' | 'bcd' 'abc' | 'pass' 'key' | 'mass' 'key' |
|---|---|---|---|---|---|---|
| **Mean** | 126.87 | 126.49 | 126.99 | 127.36 | 127.65 | 127.87 |
| **Standard deviation** | 73.670 | 73.867 | 73.376 | 73.792 | 73.769 | 73.458 |
| **Variance** | 5427.3 | 5456.4 | 5384.1 | 5445.3 | 5441.9 | 5396.1 |
| **Correlation** | -0.0044578 | | -0.013859 | | -0.0090254 | |

*Table.1 Statistical analysis (mean, standard deviation, variance) of generated vectors of 10000 pseudo-random bytes using PMSE with different password sets. The correlation between pairs of vectors is calculated below corresponding columns.*

These previous tests are completed with an image encryption and analysis using PMSE and compared with software-based random One-Time-Pad (OTP) encryption [11 -Horstmeyer et al.]. A microscopy image which includes a large parts of monochromatic dark is used (Figure 5-a). The RGB visualisation of the pseudo-random numbers generated with the same two small passwords ('aa' and 'bb') has been obtained (Figure 5-b). The image corresponds to a noisy image without any geometrical or repetition pattern that could be identified visually. The deconstructed image (Figure 5-c) is obtained with a four cases bits swapping and partial complementations (e.g. detailed in the Arduino code provided into the Supplementary data section). Some aspect of the initial image could be guessed, especially the shape of few crystals, but it is difficult to know what this image represented initially. Thanks to the partial complementations implemented, the dark part of the initial image is also affected by the "image deconstruction" process. Important parameters that could be identified into a template attack like the main color or the shape of elements are not anymore consistent using only this simple four states permutation of bits and partial complementation. The encrypted image (Figure 5-d) corresponds to the bitwise xor operation between pseudo-random matrix and deconstructed image. The identification of patterns into this image, if successful, would correspond to inconsistent informations due to the image deconstruction.

As comparison, One-Time-Pad (OTP) encryption has been also implemented for RGB images on GNU Octave (cf. code provided into supplementary data). The pseudo-random matrix (OTP) for encryption is generated using the randi() function of GNU Octave for each RGB pixel. Results of OTP and encrypted image obtained are shown in the Supplementary data section. The OTP generated with GNU Octave looks visually like truly random image without any repetition pattern (even if pure random number is ideal for fully computed numbers). The visualization of PMSE and OTP randomness calculated can be appreciated into the Supplementary data section on several encrypted images.

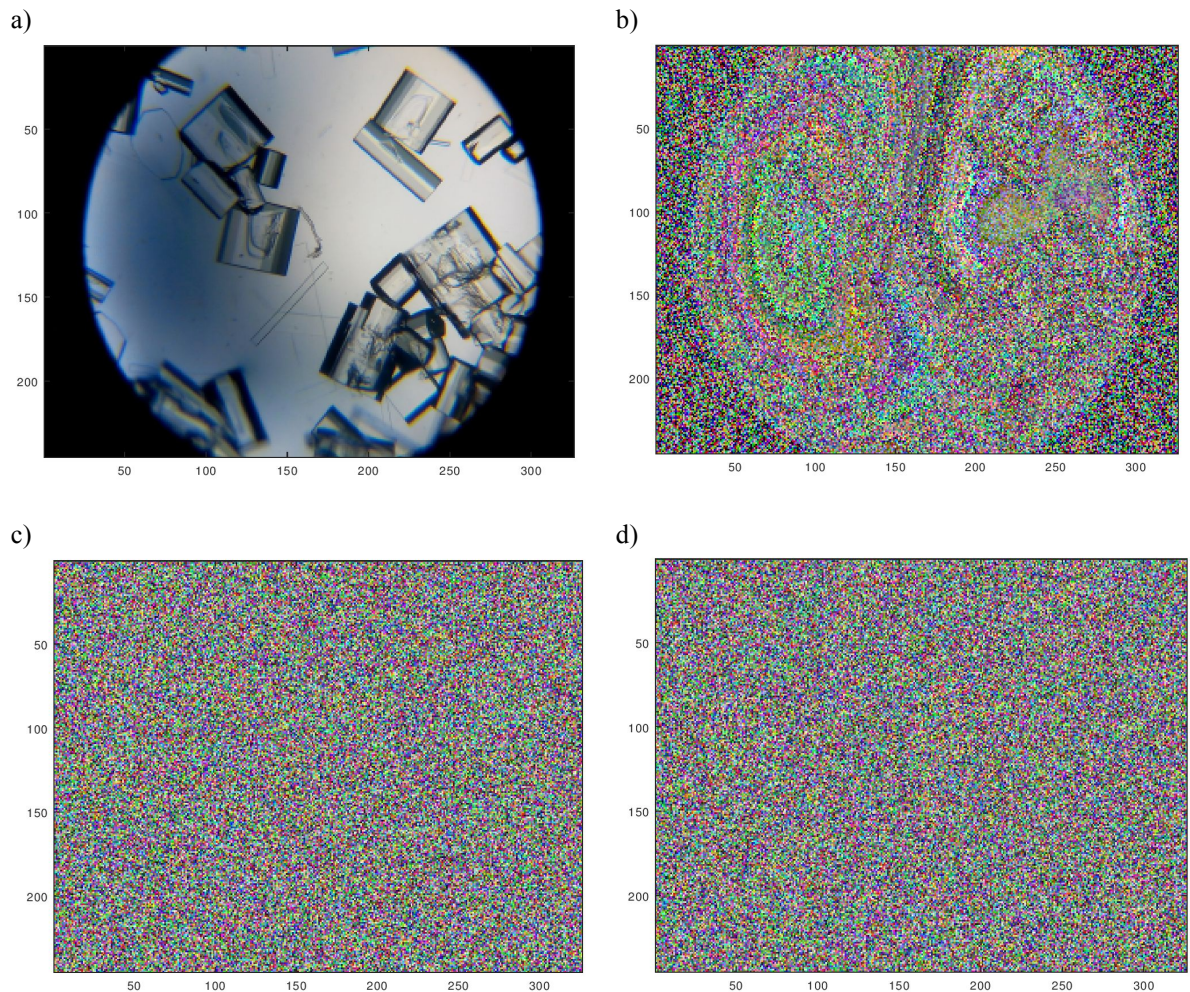

*Fig.5 a) Initial/decrypted image (microscopic view of crystals) with size of 74x98 RGB pixels, b) Image "deconstructed" with bit swapping, c) Generated pseudo-random image (matrix of 245x326x3 8 bits numbers) using PMSE, d) encrypted image using PMSE with polynomial of order 1. Images obtained with PMSE and the passwords respectively equals to 'aa' and 'bb'.*

Visualisation of the pseudo-random numbers produced is a good fast way to examine randomness in a first approach: human brain is really good at identifying redundant or ordonated patterns into an image. But statistical approach can also estimate randomness. Statistical correlation and Shannon entropy have been computed using GNU Octave on pseudo-random or encrypted images obtained with PMSE or OTP encryption. The table 2 present the correlation respectively calculated between initial image, and "deconstructed image", or pseudo-random image, or encrypted image using PMSE or OTP algorithm. We can notice that the correlation decreases if a second order polynom instead of first order is used for image deconstruction. But the decrease is not valid on final encrypted image. More generally correlation values between initial image and encrypted are below 0.002 which is already a low correlation value obtained with OPT encryption. Notice that the randomness of the OPT implemented depends on the librairie for the *randi()* function of GNU Octave. Calculated values indicate that PMSE algorithm and OTP encryptions produced the same order of correlation and entropy (more comparizon on entropy score with different images have been computed into the Supplementary data section). Therefore statistically the randomness of PMSE algorithm can be considered as strong for these image encryption examples.

| Statistical Correlation | Initial image | "Deconstructed image" | Initial image | Pseudo-random image | Initial image | Encrypted image |
|---|---|---|---|---|---|---|
| **PMSE with polynomial of order 1 (Eq.1.1)** | | | | | | |
| {'aa', 'bb'} | | 0.12394 | | 0.0020117 | | -0.0000049691 |
| {'bonjour', 'hello'} | | 0.12747 | | 0.0014874 | | 0.00086004 |
| {'mC5JLVGy6', 'tpV2gyYcK'} | | 0.12433 | | -0.0013692 | | -0.0041893 |
| **PMSE with polynomial of order 2 (Eq.1.2)** | | | | | | |
| {'mC5JLVGy6', 'tpV2gyYcK'} | | 0.10951 | | 0.0022844 | | 0.0014763 |
| **OTP** | | | | | | |
| OTP random matrix | | *NA* | | -0.0011707 | | 0.0021476 |

*Table.2 Statistical correlation calculated between the initial image and respectively the deconstructed image, the generated pseudo-random image and the encrypted image using several sets of passwords pairs of vectors is calculated below corresponding columns.*

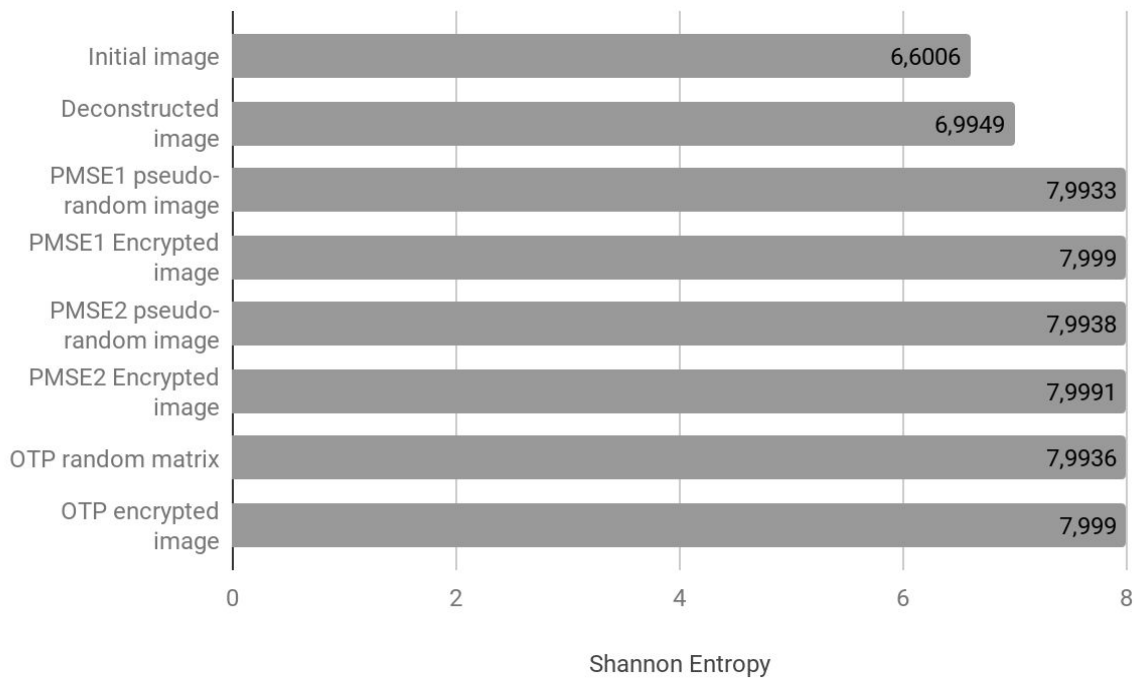

*Fig.6 Shannon entropy computed respectively on Initial image, "deconstructed" image, PMSE with 1st and 2nd order polynomial pseudo random matrix and encrypted image, and OTP random matrix obtained with randint() function (into GNU Octave) and the OPT encrypted image.*

**II.b Some comparative speed tests using Arduino**

The PMSE algorithm has been implemented in C and compiled for 8-bit microcontroller: the Arduino Uno (A*Tmega328*). The source code is provided into the Supplementary data section. Two versions have been tested, respectively with 1 or 2 password(s). The version using only one password, update the current variable $x_2$ with previous $x_1$. The version which use two passwords is the one already presented in this work and into sources codes provided. For comparison we build a program for encryption and decryption using AES. We used the open-source AES library developed by G. Gainaru for Arduino and Raspberry Pi [12]. This version implement AES in Cipher Block Chaining (CBC) mode with key of 128 or 256 bits. The comparison of processing time between AES and PMSE both running on Arduino Uno is presented in Figure 7. The test implemented encrypt and decrypt a same static text (string of 100 characters) with both algorithms. The AES and PMSE program use respectively 40% and 19% of the 2 kBytes SRAM of ATmega328 chip, so PMSE use approximately the half of dynamic memory to encrypt the same static content.

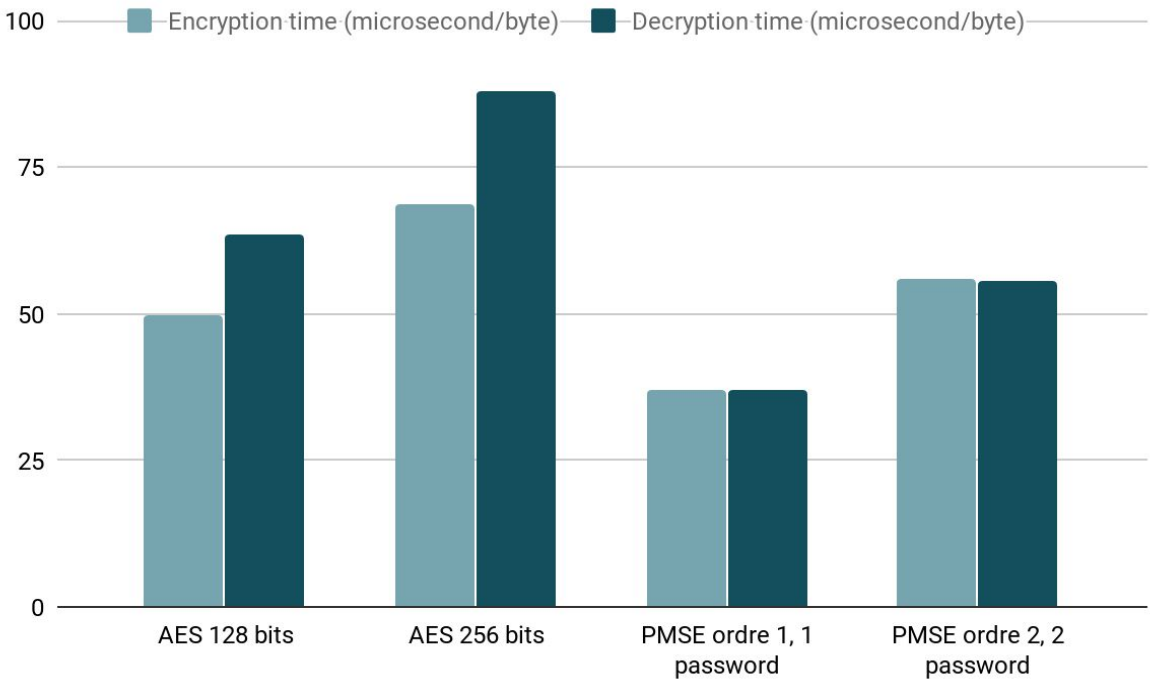

*Fig.7 Comparison of processing time on Arduino Uno between AES implementation for Arduino and PMSE using one or two passwords for encryption and decryption of 100 bytes (Clock speed: 16MHz, Architecture: Reduced Instruction Set Computer (RISC); ATmega328 processor complete one simple instruction in 63 ns).*

PMSE could have some advantage over the well established AES standard. Encryption process is done using blocks of 8-bits which is the common denominator of all processors from 8-bits microcontrollers up to 64-bits multi-cores processors. The processing time is equivalent to AES but the algorithm is compact (few lines of codes) and uses less memory (RAM and ROM). Decryption time is almost equals to encryption time. Therefore, thanks to its modularity and compactness PMSE could find application into autonomous (or specific) embedded cryptographic systems or objects. In such systems, data is stored and manipulated as encrypted by default [13] and the object contains the decryption and/or encryption mechanism. Clear data only appears in dynamic memories when decrypted and never into storage memory. Thus encryption doesn't take place only on the transmission protocol but also on stored and manipulated data. In the next section, we extend this concept to web objects and we demonstrate a self-decryptionable web encrypted object embedded into HTML page.

## III. Application : blocksnet, self-decryptionable web encrypted object

Self-decryptionable web encrypted object (WEO) is called "block" in this part. The concept of a block is described in Figure 8. Each block is web page that contains its own function to encrypt or decrypt (a part of) its content. The encryption used is the PMSE algorithm developed here for compactness and robustness. The encrypted block (or content) can be decoded using the right set of passwords. Notice that a block having an encrypted content can be encrypted again (recursively) with an additional content. Thus it is possible to build kind of "Russian doll" structure of encrypted blocks. Other structures like chain or tree of web blocks are also possible. In this case, each block contains privately one or more url which refers to other(s) block(s). Additionally checksum and timestamp are calculated for each encrypted block, so an encrypted blockchain structure can be build.

An example of blocks structure illustrating the blocksnet concept (Figure 8) has been developed and is available online. In this example, the sequence of passwords must be guessed by human brains (solve it with computers would be a challenge). The short structure of blocks takes the scenario of a small quizz game (Cf. [14]). The javascript source code or PMSE is available directly on those html blocks, illustrating the blocksnet concept. Without the right answer (or keys) to the question, the block cannot be decrypted properly and absurd ASCII characters will appear due to the improper passwords provided for decryption.

Usually standard scenario of cryptographic communication use Alice and Bob as interlocutors. In this case, block sharing between Alice and Bob could be simply done always integrating the key(s) or password(s) for the next bloc exchange into the current encrypted bloc. Only the first bloc exchange (initialization) must be done using public key cryptography or using any secured channel. This kind of exchange could be also done in broadcast mode (or master slave architecture) on a limited group of people that receive/download information. Secured and costless data sharing could be done like this: clients/users receive an encrypted email (e.g. using PGP) with initial key from server (master). Then sequence or structure of blocks can be downloaded by users and it can be sequentially decrypted on the clients/users side.

In the context of the General Data Protection Regulation (GDPR, regulation (EU) 2016/679), in Europe, such self-decryptionable web encrypted objects could be useful to encrypt some personal data (e.g. biometric or financial ones) with a modular (or personalized) encryption algorithm. Thus this kind of private cryptography based on simple web technology could be consistent at user level in order to guarantee self-privacy. With *blocksnet*, data encryption and decryption could be done on user side only, so serverless architectures are possible. Thus by using this method, no sensitive clear data are stored on servers. But data could be also easily encrypted and stored by server in order to prevent inappropriate or unconsented automatic data analysis. Blocksnet brings a dedicated response to single user data protection by implementing PMSE which is a non-standard tunable encryption algorithm. So versionnable implementations permit to encapsulate encrypted data and methods into common web pages for personalized user side decryption.

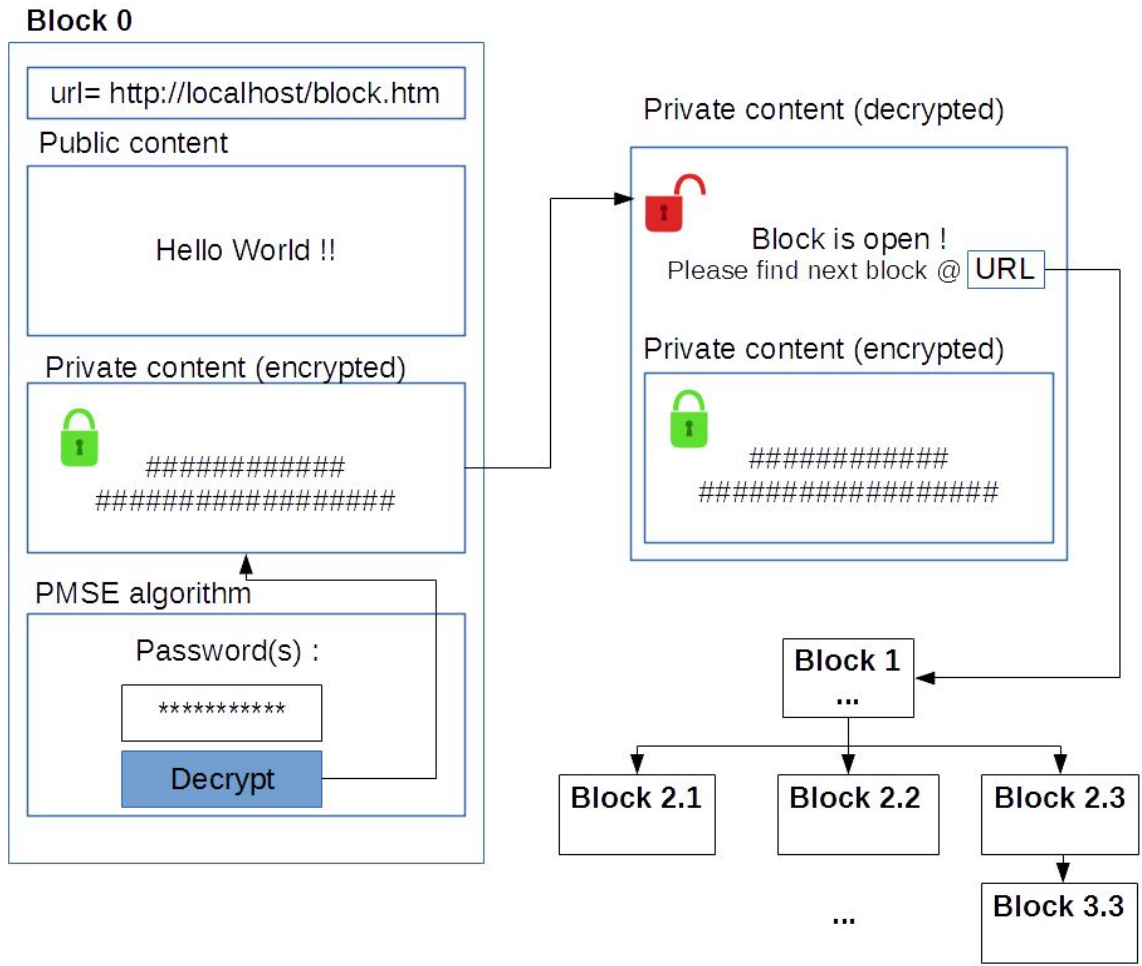

*Fig.8 Illustration of self-decryptionable web encrypted object concept. Example of potential use.*

## IV. Conclusion

The implementation of the PMSE algorithm have been detailed. The pseudo-random byte generator has been implemented and tested statistically with some parameters. Then the data deconstruction process has been presented and illustrated using image deconstruction. Finally entropy and correlation calculated on encrypted image comparing OTP and PMSE encryption gave similar results, indicating that PMSE is robust enough to be comparable to a statistically random OTP encryption.

Dedicated implementation of PMSE have been done in C and javascript. The processing performances have been compared with AES on Arduino Uno. PMSE use half less dynamic memory and about the same encryption time on this 8-bit microcontroller (ATmega328). Self-decryptionable web encrypted object implemented in javascript are also presented and illustrated. This concept will have future applications such as personalized and secured encapsulation of personal data, application, software, documents or licences.

In the future PMSE could be improved in term of compactness (less lines of code). The integration of *salt* into encrypted flux would also improved the security of encrypted data. All polynomials, parameters and versions of this modular algorithm have not been tested yet. But the modularity of the algorithm itself allows to change the cryptographic core at each use. Therefore, the difficulty to identify version and polynomial used for each encryption is improving the effective security of the algorithm.

# Supplementary data

### Illustration of PMSE entropy with images encryption :

The following images (cf. Table 1) are produced for several PMSE encryptions tests with the two same passwords (pass1='PMSE_encryption' and pass2 = 'blocksnet'). The comparison is done with One-Time-Pad (OTP) encryption (cf. GNU Octave code : One Time Pad (OTP) image encryption). Random pads (OPT) are computed with the function *randi()* of Octave. The Shannon entropy has been calculated with Octave for each image. It appears that for every encryption performed, the PMSE encryption give better or equal entropy score than the OTP encryption performed. Therefore the PMSE algorithm seems to be as good as an OTP encryptions using computed "random" masks.

Table 1 - Comparison on image encryptions using PMSE and OTP

| Initial image | PMSE encryption | | OTP encryption |
|---|---|---|---|
| Image entropy = 4.1333 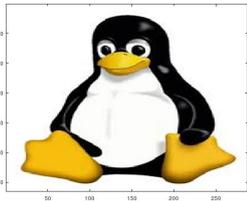 | Deconstructed Image entropy =5.0029 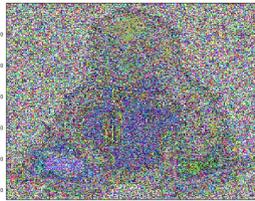 | Encrypted Image entropy = **7.9989** 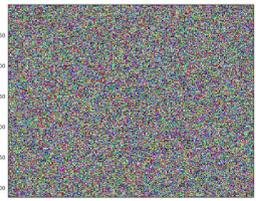 | Encrypted Image entropy = **7.9985** 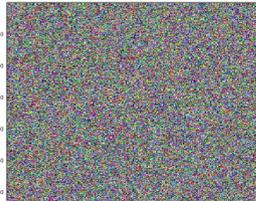 |
| Image entropy = 7.5434 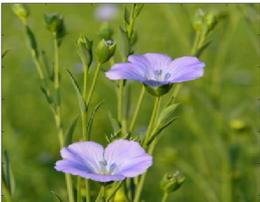 | Deconstructed Image entropy = 7.8807 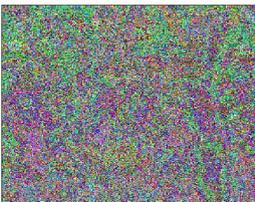 | Encrypted Image entropy = **7.9994** 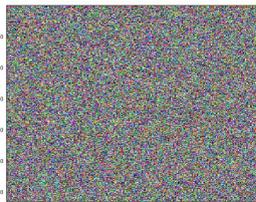 | Encrypted Image entropy = **7.9993** 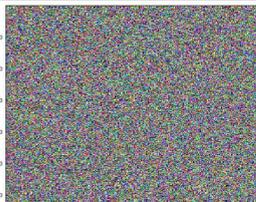 |
| Image entropy = 7.3110 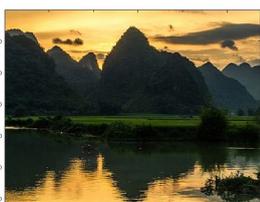 | Deconstructed Image entropy = 7.7297 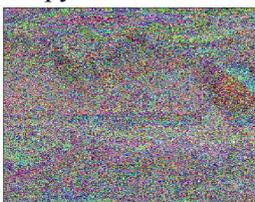 | Encrypted Image entropy = **7.9994** 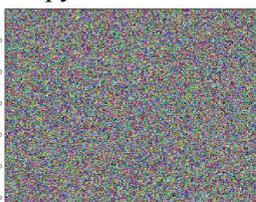 | Encrypted Image entropy = **7.9994** 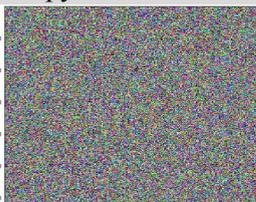 |

| Image entropy = 5.6249 | Deconstructed Image entropy = 6.2439 | Encrypted Image entropy = **7.9991** | Encrypted Image entropy = **7.9988** |
|---|---|---|---|
| 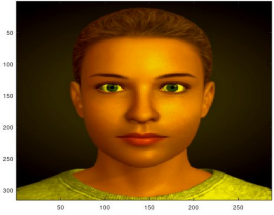 | 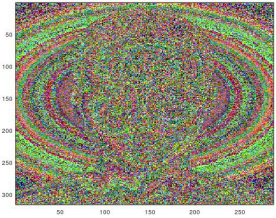 | 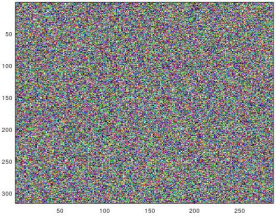 | 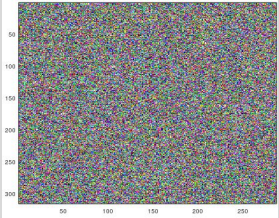 |

**GNU Octave code : PMSE image encryption and decryption**

```
%%%%%%%%%%%%%%%%%%%%%%%%%%%%%%%%%%%%%%%%%%%%%%%%%%%
%% PMSE image encryption and decryption - 10/2018    %%
%% This code has been developed by Etienne LEMAIRE   %%
%%%%%%%%%%%%%%%%%%%%%%%%%%%%%%%%%%%%%%%%%%%%%%%%%%%

clear all
close all
pkg load image

pass1='PMSE_encryption'
pass2 = 'blocksnet';

lp1  = sizeof(pass1);
lp2  = sizeof(pass2);

%iMG = imread('TUX_TEST.bmp');
iMG = imread('GRASS_TEST.bmp');
%iMG = imread('LAC_TEST.bmp');
%iMG = imread('PORTRAIT_TEST.bmp');
gen_key = iMG;
gen_key = gen_key - iMG;

[xl yl zl] = size(iMG);

figure()
imagesc(iMG);
title('Initial/Decrypted image');

%%%%%%%%%%%%%%%%%%%%%%%%%%%%%
%% Image encryption using PMSE
a1 = 77;
b1 = 51;
c1 = 13;
xc = 67;
x3 = 11;
xt = 234;
x0 = 88;
x1 = 97;
x2 = 132;
%dt_ = 11;
tmp1 = 1;
tmp2 = 1;
x_cs = 10;
Yn = 9;

for i=1:xl
  for j=1:yl
      for k=1:zl
```

```
        I = i+j+k;

        %% Polynomial
        Yn = x2*(I*I) + x1*I + Yn/4;

        xa = bitand(Yn,4278190080)/(2^24) ; % (Yn & 0xFF000000)>>24 ;
        xb = bitand(Yn,16711680)/(2^16); % (Yn & 0xFF0000)>>16 ;
        xc = bitand(Yn,65280)/(2^8); % (Yn & 0xFF00)>>8 ;
        xd = bitand(Yn, 255); % Yn & 0xFF;
        x0 = bitxor(xd, xc) + bitxor(xa,xb); % x0 = xd^xc + xa^xb ;

        x1 = double(pass1(1+ mod(I,lp1-1)));
        x2 = double(pass2(1+ mod(1 + x1 +I,lp2-1)));
        x3 = mod(( I*x1 - x3*x2),255);

        x1 =  bitxor(x0,x1);
        tmp1 = bitxor(x1,x2);
        tmp2 = bitxor(x3,xt);
        xt =  bitand(bitxor(tmp1,tmp2),255); % xt = xt ^ x0 ^ (x1 ^ x2) ^ x3 ;

        if (xt == 0)
        a1 = mod(I,31);
        b1 = mod(I,23);
        c1 = mod(I,57);
        x3 = mod(I,233);
        xt = mod(I,157);
        x0 = mod(I,103);
        x1 = mod(I,97);
        x2 = mod(I,131);
        end;

        % gen key storage
        gen_key(i,j,k) = xt;

        % DATA encoding : bits switching
        dat_ = iMG(i,j,k);
        dat_0xC0 = bitand(dat_,192);
        dat_0x30 = bitand(dat_,48);
        dat_0x0C = bitand(dat_,12);
        dat_0x03 = bitand(dat_,3);
        dat_0x0F = bitand(dat_,15);
        dat_0xF0 = bitand(dat_,240);

        %xn = x1+x2;
        xn = x0;

        if ( mod(xn,4)==0)
        dat_r = (dat_0x0F*(2^4)) + (dat_0xF0/(2^4));
        dat_r = bitxor(dat_r, 192); % dat_r^0xC0
        end;

        if ( mod(xn,4)==1)
        dat_r = (dat_0x0F*(2^4)) + (dat_0xF0/(2^4));
        dat_r = bitxor(dat_r, 12); end;
        if ( mod(xn,4)==2)
        dat_r = dat_0x0C*(2^2) + dat_0x03*(2^2) + dat_0xC0/(2^6) + dat_0x30*(2^2);
        dat_r = bitxor(dat_r, 192); end;

        if ( mod(xn,4)==3)
        dat_r = dat_0x0C*(2^2) + dat_0x03*(2^6) + dat_0xC0/(2^6) + dat_0x30/(2^2);
        dat_r = bitxor(dat_r, 12); end;
```

```
        % "randomized" data storage
        rand_data(i,j,k) = dat_r;
        
        % symmetric XOR (en)coding
        %iMG_crypt(i,j,k) = bitxor(iMG(i,j,k),xt);
        iMG_crypt(i,j,k) = bitxor(dat_r,xt);
        
        % checksum
        x_cs = bitxor(iMG_crypt(i,j,k), x_cs) + x_cs ;
        
        end;
   end;
  end;
E_key = entropy(gen_key)
E_img_enc = entropy(iMG_crypt)
E_dimg = entropy(rand_data)
E_img_ = entropy(iMG)

  %moy_ = mean(res)
  ecart_type_img = std(iMG(:))
  ecart_type_key = std(gen_key(:))
  ecart_type_dec_img = std(rand_data(:))
  ecart_type_enc_img = std(iMG_crypt(:))
  %variance_ = var(res)
  autocorrelation = cov(iMG(:),iMG(:))./(ecart_type_img.*ecart_type_img)
  correlation_ik = cov(iMG(:),gen_key(:))./(ecart_type_img.*ecart_type_key)
  correlation_id = cov(iMG(:),rand_data(:))./(ecart_type_img.*ecart_type_dec_img)
  correlation_ie = cov(iMG(:),iMG_crypt(:))./(ecart_type_img.*ecart_type_dec_img)

figure
imagesc(gen_key);
title('Generated pseudo-random flux')

figure
imagesc(iMG_crypt);
title('Encoded image')

figure
imagesc(rand_data);
title('Deconstructed image')

%%%%%%%%%%%%%%%%%%%%%%%%%%%%%%%%%%%%%%%%%%%%%
%% Image decoding using PMSE

a1 = 77;
b1 = 51;
c1 = 13;
xc = 67;
x3 = 11;
xt = 234;
x0 = 88;
x1 = 97;
x2 = 132;
%dt_ = 11;
tmp1 = 1;
tmp2 = 1;
x_cs = 10;
Yn = 9;

for i=1:xl
  for j=1:yl
      for k=1:zl
      
      I = i+j+k;
```

```matlab
        %% Polynomial
        Yn = x2*(I*I) + x1*I + Yn/4;

        xa = bitand(Yn,4278190080)/(2^24) ; % (Yn & 0xFF000000)>>24 ;
        xb = bitand(Yn,16711680)/(2^16); % (Yn & 0xFF0000)>>16 ;
        xc = bitand(Yn,65280)/(2^8); % (Yn & 0xFF00)>>8 ;
        xd = bitand(Yn, 255); % Yn & 0xFF;
        x0 = bitxor(xd, xc) + bitxor(xa,xb); % x0 = xd^xc + xa^xb ;

        x1 = double(pass1(1+ mod(I,lp1-1)));
        x2 = double(pass2(1+ mod(1 + x1 +I,lp2-1))); % char selection for pseudo
rand xd
        x3 = mod(( I*x1 - x3*x2),255);

        x1 =  bitxor(x0,x1);
        tmp1 = bitxor(x1,x2);
        tmp2 = bitxor(x3,xt);
        xt =  bitand(bitxor(tmp1,tmp2),255); % xt = xt ^ x0 ^ (x1 ^ x2) ^ x3 ;

        if (xt == 0)
        a1 = mod(I,31);
        b1 = mod(I,23);
        c1 = mod(I,57);
        x3 = mod(I,233);
        xt = mod(I,157);
        x0 = mod(I,103);
        x1 = mod(I,97);
        x2 = mod(I,131);
        end;

        % gen key storage
        gen_key(i,j,k) = xt;

        % symmetric XOR (de)coding
        iMG_dec(i,j,k) = bitxor(iMG_crypt(i,j,k),xt);

        %xn = x1+x2;
        xn = x0;

        dat_r = iMG_dec(i,j,k);

        if ( mod(xn,4)==0)
        dat_r = bitxor(dat_r, 192); % dat_r^0xC0

        end;

        if ( mod(xn,4)==1)
        dat_r = bitxor(dat_r, 12); % dat_r^0x0C

        end;
        if ( mod(xn,4)==2)
        dat_r = bitxor(dat_r, 192); % dat_r^0xC0

        end;

        if ( mod(xn,4)==3)
        dat_r = bitxor(dat_r, 12); % dat_r^0x0C

        end;

        % DATA decoding : bits switching
        dat_0xC0 = bitand(dat_r,192);
```

```
        dat_0x30 = bitand(dat_r,48);
        dat_0x0C = bitand(dat_r,12);
        dat_0x03 = bitand(dat_r,3);
        dat_0x0F = bitand(dat_r,15);
        dat_0xF0 = bitand(dat_r,240);

        if ( mod(xn,4)==0)
        dat_r = (dat_0x0F*(2^4)) + (dat_0xF0/(2^4));

        end;

        if ( mod(xn,4)==1)
        dat_r = (dat_0x0F*(2^4)) + (dat_0xF0/(2^4));
        end;
        if ( mod(xn,4)==2)
        dat_r = dat_0x0C/(2^2) + dat_0x03*(2^6) + dat_0xC0/(2^2) + dat_0x30/(2^2);
        end;

        if ( mod(xn,4)==3)
        dat_r = dat_0x0C*(2^2) + dat_0x03*(2^6) + dat_0xC0/(2^6) + dat_0x30/(2^2);
        end;

        % recontructed data
        iMG_dec2(i,j,k) = dat_r;

        % checksum
        x_cs = bitxor(iMG_crypt(i,j,k), x_cs) + x_cs ;

        end;
   end;
  end;
figure
imagesc(iMG_dec);
title('Partially decoded image')

figure
imagesc(iMG_dec2);
title('Reconstructed image')
```

# Illustration of PMSE pseudo-random generated images with several sets of passwords :

Images of the pseudo-random numbers produced with PMSE and several set of passwords, the corresponding Shannon entropy for each image is calculated.

**Initial image**
Image entropy = 6.6006

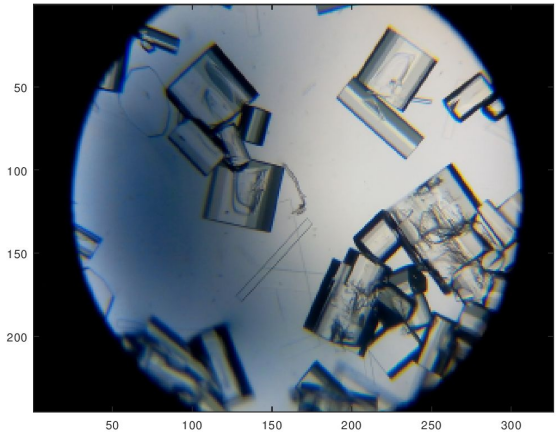

**"Deconstructed" image**
Image entropy = 6.9949

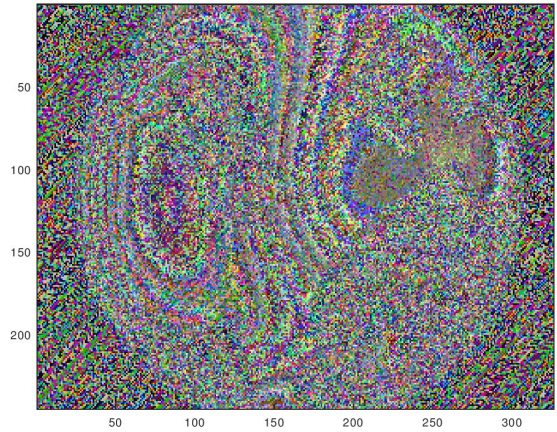

**1st order non-recursive polynomial : $Y_n = x_2*i + x_1$**

**Passwords: {'aa', 'bb'}**
Pseudo-random image entropy = 7.9933

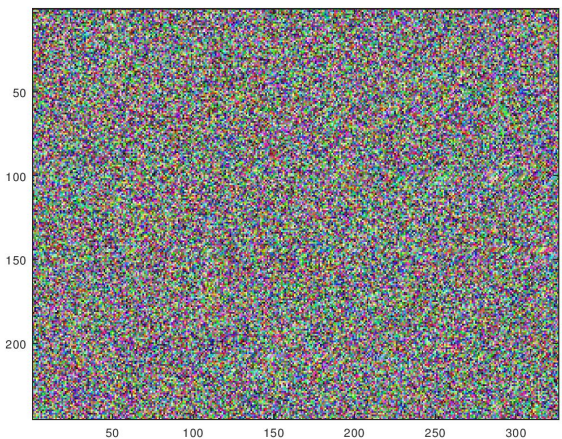

**Passwords: {'bonjour', 'hello'}**
Pseudo-random image entropy = 7.9929

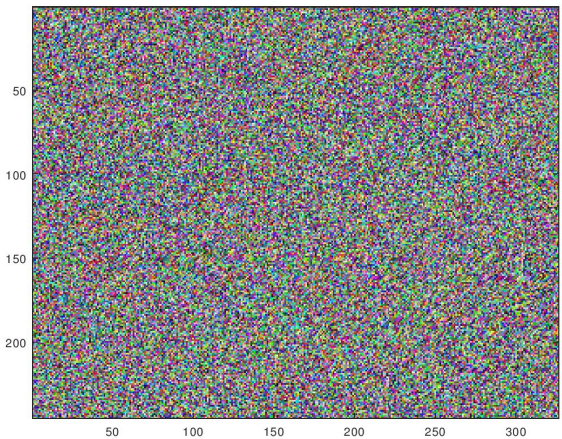

**Passwords: {'dA4wJ6', 'lUVmX2'}**
Pseudo-random image entropy = 7.9927

**Passwords: {'mC5JLVGy6', 'tpV2gyYcK'}**
Pseudo-random image entropy = 7.9929

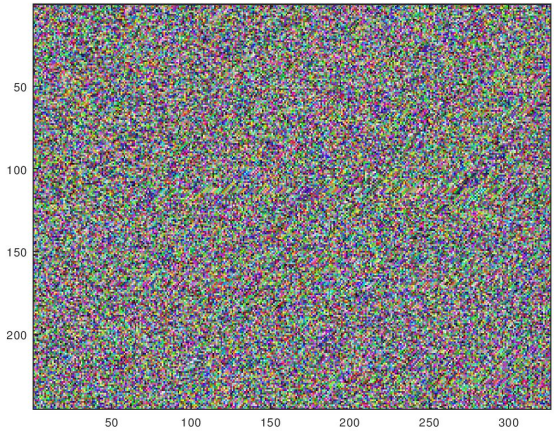
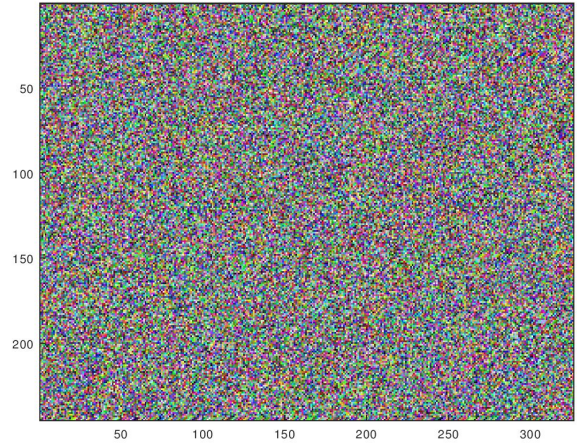

## 2nd order recursive polynomial: $Y_n = x_2 \ast i^2 + x_1 \ast i + Y_{n-1}/4$

**Passwords: {'aa', 'bb'}**
Pseudo-random image entropy = 7.9938

**Passwords: {'bonjour', 'hello'}**
Pseudo-random image entropy = 7.9939

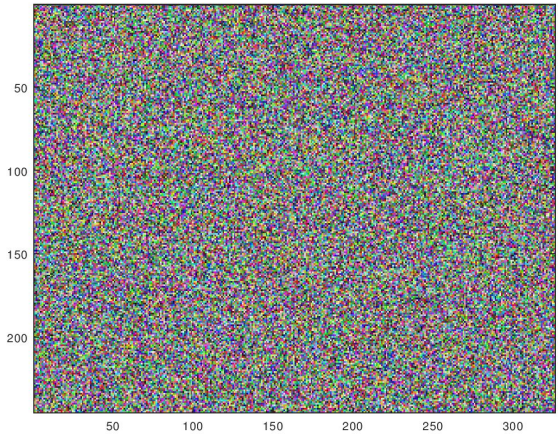
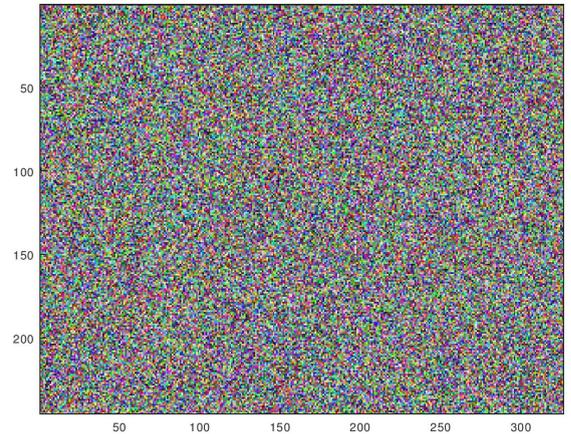

**Passwords: {'dA4wJ6', 'lUVmX2'}**
Pseudo-random image entropy = 7.9936

**Passwords: {'mC5JLVGy6', 'tpV2gyYcK'}**
Pseudo-random image entropy = 7.9939

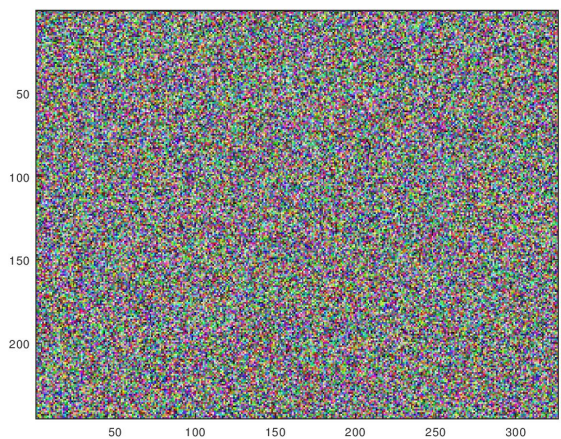
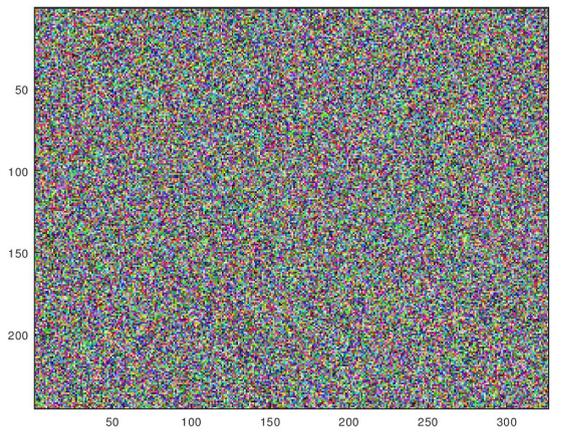

NB: the entropy of images with differents set of passwords has been computed using GNU Octave : https://octave.sourceforge.io/image/function/entropy.html

# GNU Octave code : pseudo-random flux generator from two passwords

```
%%%%%%%%%%%%%%%%%%%%%%%%%%%%%%%%%%%%%%%%%%%%%%
%% Pseudo-random byte generator of PMSE - 10/2018    %%
%% This code has been developed by Etienne LEMAIRE   %%
%%%%%%%%%%%%%%%%%%%%%%%%%%%%%%%%%%%%%%%%%%%%%%

clear all
close all

pass1 = 'aa';
pass2 = 'bb';

lp1  = sizeof(pass1);
lp2  = sizeof(pass2);
Long_msg = 10000;
x3 = 11;
xt = 234;
x0 = 88;
x1 = 77;
x2 = 132;
res_srt = '';

for i=1:Long_msg

      Yn = x2*(i) + x1;
      xa = bitand(Yn/(2^24), 255);
      xb = bitand(Yn/(2^16), 255); % (xa & 0xFF0000)>>16 ;
      xc = bitand(Yn/(2^8),255); % (xa & 0xFF00)>>8 ;
      xd = bitand(Yn, 255); % xa & 0xFF;
      x0 = bitxor(xd, xc) + bitxor(xa, xb); % x0 = xd ^ xc + xb ;

  x1 = double(pass1(1+ mod(i,lp1-1)));
  x2 = double(pass2(1+ mod(i+ x1,lp2-1))); % char selection for pseudo rand x0
  x3 = mod((i + x3 + x2 - x1),255);

  x1 =  bitxor(x0,x1);
  tmp1 = bitxor(x1,x2);
  tmp2 = bitxor(x3,xt);
  %xt = bitxor(tmp1,tmp2);
  xt =  bitand(bitxor(tmp1,tmp2),255);
  if (xt == 0)
      x3 = mod(i,233);
      xt = mod(i,157);
      x0 = mod(i,103);
      x1 = mod(i,97);
      x2 = mod(i,131);
      end;
  res(i) = xt;
  res_srt = strcat(res_srt,char(xt));
  end;

  % Stats
  moy_ = mean(res)
  ecart_type = std(res)
  variance_ = var(res)
  correlation = cov(res,res_)./(ecart_type.*ecart_type_)

  % FFT de res(i)
  Y = fft(res);
  [u L]=size(res);
  P2 = abs(Y/L);
  P1 = P2(1:L/2+1);
```

```
P1(2:end-1) = 2*P1(2:end-1);
figure()
hold on
Fs =1;
f = Fs*(0:(L/2))/L;
semilogy(f,P1) %(1:5001)
%title('Single-Sided Amplitude Spectrum of X(t)')
xlabel('f')
ylabel('|H(f)|')

%disp(res_srt);
%figure
%plot(res, '.b')
%xlabel('Iterations (length)')
%ylabel('Ascii code [0-255]')
figure
plot(res(1:512), '.-')
xlabel('Iterations (length)')
ylabel('Ascii code [0-255]')
figure
plot(res(4096:(4096+512)), '.-')
xlabel('Iterations (length)')
ylabel('Ascii code [0-255]')
figure
hist(res, 255)
xlabel('Ascii code [0-255]')
ylabel('Number of occurrences')
%histfit(res)
```

**One Time Pad (OTP) encryption figures:**

Images produced with the OTP encryption (cf. Octave code provided below), the corresponding Shannon entropy for each image is calculated.

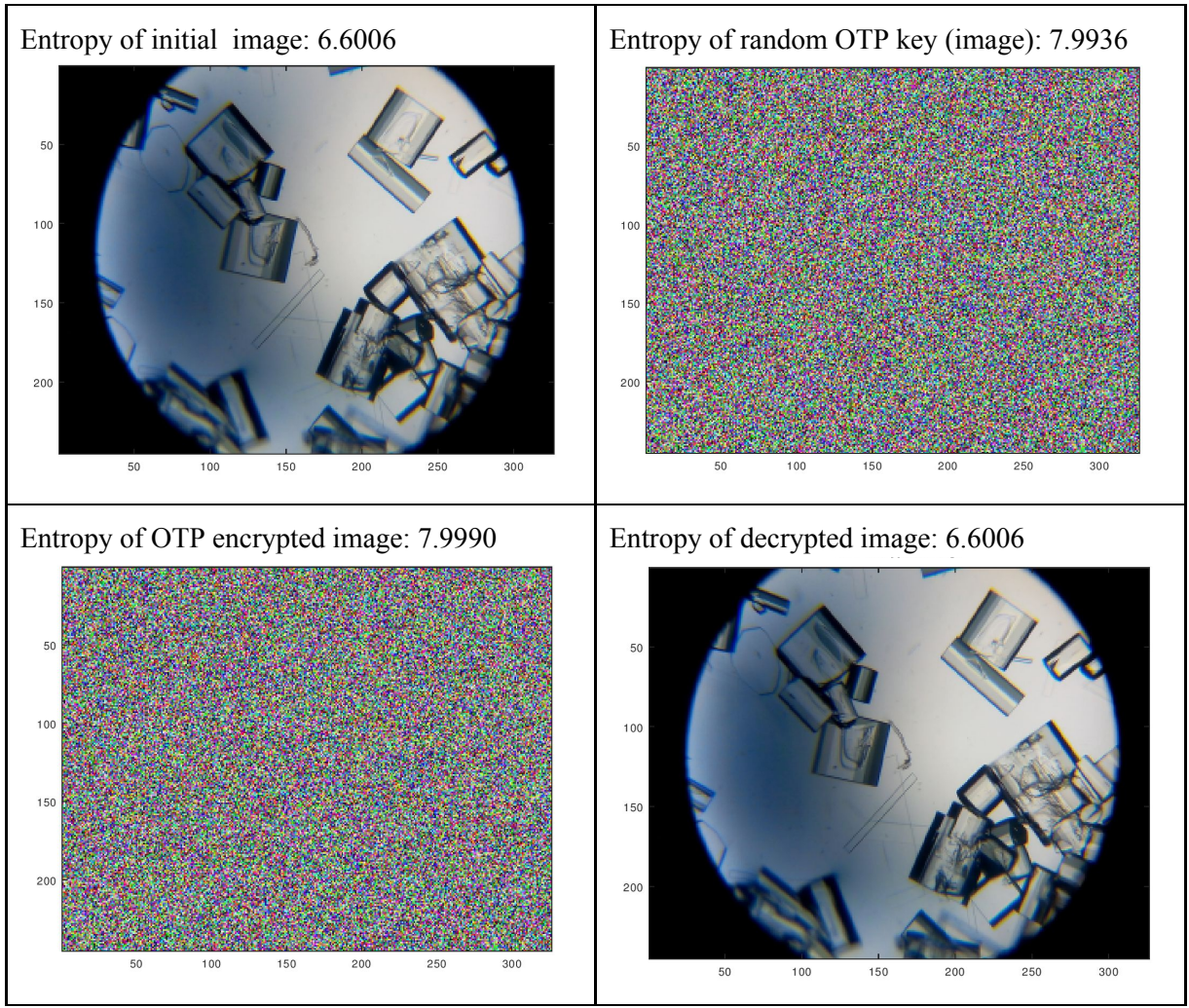

**GNU Octave code : One Time Pad (OTP) image encryption**

```
%%%%%%%%%%%%%%%%%%%%%%%%%%%%%%%%%%%%%%%%%%%%%%%%%%%%
%% One Time Pad image encryption test - 10/2018   %%
%% This code has been developed by Etienne LEMAIRE %%
%%%%%%%%%%%%%%%%%%%%%%%%%%%%%%%%%%%%%%%%%%%%%%%%%%%

clear all
close all
%pkg load statistics
pkg load image
pkg load communications
```

```
% load image
iMG = imread('IMG_TEST.bmp');
gen_key = iMG;
gen_key = gen_key - iMG;

figure()
imagesc(iMG);
title('Initial image');

% compute image entropy
E_img_ = entropy(iMG)

[x y z]= size(iMG);

% One Time Pad (OTP) generation using randint
% https://octave.sourceforge.io/communications/function/randint.html
R_otp = randint (x, y, 255);
G_otp = randint (x, y, 255);
B_otp = randint (x, y, 255);

key_otp = iMG - iMG;
key_otp(:,:,1) = R_otp;
key_otp(:,:,2) = G_otp;
key_otp(:,:,3) = B_otp;

% display OTP random matrix
figure()
imagesc(key_otp);
title('OTP key');

% compute OTP random key entropy
E_key_ = entropy(key_otp)
% correlation between random key and encrypted image
correlation_key_img =
cov(iMG(:),key_otp(:))./(std(iMG(:)).*std(key_otp(:)))

% One Time Pad (OTP) encryption
iMG_otp = bitxor(iMG, key_otp);

figure()
imagesc(iMG_otp);
title('OTP encrypted image');

% compute encrypted image entropy
E_enc_ = entropy(iMG_otp)
% correlation between initial image and encrypted image
correlation_enc_img =
cov(iMG(:),iMG_otp(:))./(std(iMG(:)).*std(iMG_otp(:)))

% One Time Pad (OTP) decryption
iMG_dec = bitxor(iMG_otp, key_otp);

figure()
imagesc(iMG_dec);
title('OTP decrypted image');
```

**Arduino code : PMSE on Arduino**

```
//////////////////////////////////////////////////////////
// PMSE de/encryption implemented on Arduino Uno - 10/2018   //
// This code has been developed by Etienne LEMAIRE           //
//////////////////////////////////////////////////////////

char texte[100] = "Bonjour, je teste PMSE, pretty modular symetric
encryption, ceci est simplement un test! Encore un!";
char pass_1[14] = "zc2dhvnepc#b91";
char pass_2[8] = "coucou21";
char iv_[24] = "1q23df5r8tyb6d9r5t7k6s4e";

void setup() {
  // Serial setup
  Serial.begin(9600);

}

void loop() {
  //Encryption time
  Serial.print("- encryption time [us]: ");
  unsigned long ms = micros ();
  pmse_encrypt(texte, 100, pass_1, 14, pass_2, 8, iv_);
  Serial.println(micros() - ms);
  Serial.println(texte);
  //Decryption time
  Serial.print("- decryption time [us]: ");
  ms = micros ();
  pmse_decrypt(texte, 100, pass_1, 14, pass_2, 8, iv_);
  Serial.println(micros() - ms);
  Serial.println(texte);
}

///////////////////////////////////
// PMSE Library v0-beta 10/18    //
// developed by Etienne LEMAIRE  //
///////////////////////////////////

// encryption fonction
void pmse_encrypt(char *msg, int l_msg, char *pass, int l_pass, char
*pass2, int l_pass2, char iv[24]){

int x0 = (int)iv[0];
int x1 = (int)iv[1];
```

```c
int x2 = (int)iv[2];
int x3 = (int)iv[3];

int a1 = (int)iv[4];
int b1 = (int)iv[5];
int c1 = (int)iv[6];

char xt = iv[7], data =0, xa=0, xb=0, xc=0, xd=0;

int tmp1 = (int)iv[8];
int tmp2 = (int)iv[9];
int Yn = (int)iv[10];

int i=0;

	for ( i=0; i<l_msg ; i++){

	// Pseudo random byte generation
	Yn = x2*i + x1;

	xa = (Yn & 0xFF000000)>>24 ;
	xb = (Yn & 0xFF0000)>>16 ;
	xc = (Yn & 0xFF00)>>8 ;
	xd = Yn & 0xFF;
	x0 = (xa^xb^xc^xd);

	x1 = pass[i % l_pass];
	x2 = pass2[(i+x1)%(l_pass2)];
	x3 = (i*x1 - x3*x2)%255;
	xt = (xt^x0^x1^x2^x3)&0xFF;

	if (xt==0){
		a1 = i%iv[11];
		b1 = i%iv[12];
		c1 = i%iv[13];
		xt = i%iv[14];
		x0 = i%iv[15];
		x1 = i%iv[16];
		x2 = i%iv[17];
		x3 = i%iv[18];

	}

	// data "desconstruction"
	data = msg[i];
	if ((xd & 0x03)==0){
		data = ((data&0x0F)<<4) + ((data&0xF0)>>4);
		// data = data^0xC0;
	}
	if ((xd & 0x03)==1){
		data = ((data&0x3F)<<2) + ((data&0xC0)>>6);
		// data = data^0x0A;
	}
	if ((xd & 0x03)==2){
		data = ((data&0x33)<<2) + ((data&0xCC)>>2);
		// data = data^0xA0;
	}
```

```c
        if ((xd & 0x03)==3){
            data = ((data&0x1F)<<3) + ((data&0xE0)>>5);
            // data = data^0x0C;
        }

        // data encryption
        msg[i] = data^xt;

        }

}

// decoding fonction
void pmse_decrypt(char *msg, int l_msg, char *pass, int l_pass, char
*pass2, int l_pass2, char iv[24]){

int x0 = (int)iv[0];
int x1 = (int)iv[1];
int x2 = (int)iv[2];
int x3 = (int)iv[3];

int a1 = (int)iv[4];
int b1 = (int)iv[5];
int c1 = (int)iv[6];

char xt = iv[7], data =0, xa=0, xb=0, xc=0, xd=0;

int tmp1 = (int)iv[8];
int tmp2 = (int)iv[9];
int Yn = (int)iv[10];

int i=0;

    for ( i=0; i<l_msg ; i++){

    // Pseudo random byte generation
    Yn = x2*i + x1;

    xa = (Yn & 0xFF000000)>>24 ;
    xb = (Yn & 0xFF0000)>>16 ;
    xc = (Yn & 0xFF00)>>8 ;
    xd = Yn & 0xFF;
    x0 = (xa^xb^xc^xd);

    x1 = pass[i % l_pass];
    x2 = pass2[(i+x1)%(l_pass2)];
    x3 = (i*x1 - x3*x2)%255;
    xt = (xt^x0^x1^x2^x3)&0xFF;

    if (xt==0){
        a1 = i%iv[11];
        b1 = i%iv[12];
        c1 = i%iv[13];
        xt = i%iv[14];
        x0 = i%iv[15];
        x1 = i%iv[16];
```

```
                x2 = i%iv[17];
                x3 = i%iv[18];

        }
        data = msg[i];

        // partial decoding
        data = data^xt;

                // data "reconstruction"

        if ((xd & 0x03)==0){
                data = ((data&0x0F)<<4) + ((data&0xF0)>>4);
                // data = data^0xC0;
        }
        if ((xd & 0x03)==1){
                data = ((data&0xFC)>>2) + ((data&0x03)<<6);
                // data = data^0x0A;
        }
        if ((xd & 0x03)==2){
                data = ((data&0x33)<<2) + ((data&0xCC)>>2);
                // data = data^0xA0;
        }
        if ((xd & 0x03)==3){
                data = ((data&0xF8)>>3) + ((data&0x07)<<5);
                // data = data^0x0C;
        }

        msg[i]=data;
        }

}
```